\documentclass[aps,prl,twocolumn,showpacs]{revtex4}
\usepackage{amsmath,graphicx,amsbsy,amssymb,epsfig,latexsym,wasysym,pifont,mathrsfs} 
\usepackage{float}
\newfloat{widefig}{thp}{lop}
\begin{document}

\title{Discontinuous transition in an equilibrium percolation model
  with suppression}

\author{B. Roy}%
\author{S. B. Santra}%
\email{santra@iitg.ernet.in}
\affiliation{Department of Physics, Indian Institute of Technology
Guwahati, Guwahati-781039, Assam, India.}

\date{\today}
  
\begin{abstract}
  Discontinuous transition is observed in the equilibrium cluster
  properties of a percolation model with suppressed cluster growth as
  the growth parameter $g_0$ is tuned to the critical threshold at
  sufficiently low initial seed concentration $\rho$ in contrast to
  the previously reported results on non-equilibrium growth models. In
  the present model, the growth process follows all the criteria of
  the original percolation model except continuously updated
  occupation probability of the lattice sites that suppresses the
  growth of a cluster according to its size.  As $\rho$ varied from
  higher values to smaller values, a line of continuous transition
  points encounters a coexistence region of spanning and non-spanning
  large clusters. At sufficiently small values of $\rho$ ($\le 0.05$),
  the growth parameter $g_0$ exceeds the usual percolation threshold
  and generates compact spanning clusters leading to discontinuous
  transitions.
\end{abstract}

\maketitle 

Recently, percolation transition (PT) has been reported as a first
order discontinuous transition in a model of explosive percolation
(EP) by Achlioptas {\em et al} \cite{AP}. However, percolation is well
known as a model of second order continuous phase transition (CPT) and
widely applied in a variety of problems ranging from sol-gel to metal
insulator transition \cite{stauffer,bunde,grimmett,sahimi}. Instead of
the original equilibrium percolation model, a series of
non-equilibrium growth models \cite{epj223} then proposed to
demonstrate first order PT. In these models, imposing the product (or
sum) rule to occupy a bond in growing a cluster, a discontinuous jump
in the size of the largest cluster at a delayed percolation threshold
is characterized as discontinuous phase transition (DPT). However,
soon a controversy that the EP is a DPT or not erupts on the basis of
slow convergence of asymptotic cluster properties in the
$L\rightarrow\infty$ limit \cite{costa,rio,fried}. For example, some
of the Euclidean lattice models \cite{ziff,arg2} of EP were found
inconclusive in their nature of transition \cite{grass,unsul}. In the
spanning cluster avoiding (SCA) model of EP, it was claimed that there
exists an upper critical dimension below which the transitions will be
discontinuous \cite{avd} though in this model the transition occurs at
unit probability, a trivial percolation threshold. A few growth models
\cite{kim,herr1,herr2,manna,bfwm} in Euclidean space, however, are
found to display first order DPT. It seems beside CPT and true first
order DPT there exists a mixed DPT in which characteristics of both
first order and second order transitions appear \cite{tric,radi} and
the system possess unusual finite size scaling (FSS) \cite{hyb}. In
most of the cases, except the jump in the order parameter the other
aspects of first order transition such as phase co-existence,
nucleation, etc.  are ignored \cite{grass,janssen}. More importantly,
not only the understanding of the origin of DPT remains incomplete but
also it is not yet demonstrated in the context of equilibrium
percolation model.

In this letter, we propose a two parameter equilibrium percolation
model keeping nucelation and growth as the main ingredient. The
parameters are the initial seed concentration $\rho$ and a growth
parameter $g_0$. The model displays CPT, mixed DPT and finally true
first order DPT at suitable range of parameter values. The DPT in this
model is not only characterized by the jump in the order parameter but
also supported by the presence of phase co-existence. The model not
only distinguishes clearly the features of different PTs but also
captures most of the essential features of several different EP
models.

The model is developed on a $2$-dimensional ($2$d) square lattice of
size $L\times L$ occupying the lattice sites randomly with an initial
seed concentration $\rho$. Clusters of occupied sites, connected by
nearest neighbor (NN) bonds, are formed. The initial cluster size
distribution is determined identifying the clusters by Hoshen-Kopelman
algorithm \cite{HK}. The clusters are then arranged in an ascending
order according to their sizes $s$. These finite clusters are then
grown sequentially starting from the smallest cluster with a size
dependent probability. At a Monte Carlo (MC) time step $t$, the growth
probability $g_s(t)$ of a cluster of size $s$ is given by
\begin{equation}
\label{invf}
g_s(t)=g_0 \exp\left[-\{s(t)-1\}/s_{\rm large}(t)\right]
\end{equation}
where the growth parameter $g_0$ is a constant between $[0,1]$ and
$s_{\rm large}(t)$ is the size of the largest cluster present at that
time. At any time $t$, the value of $g_s(t)$ is the smallest ($g_0/e$)
for the largest cluster and it is largest ($g_0$) for the smallest
cluster ($s=1$). Accordingly, the model is called suppressed cluster
growth percolation (SCGP) which is quite different from the controlled
largest cluster growth model of EP \cite{herr1}. In a single MC step,
only a single layer of empty NN perimeter (both internal and external)
sites of a cluster are occupied with its growth probability
$g_s(t)$. Once a site is rejected with probability $(1-g_s(t))$, the
site remains unoccupied throughout the growth process as in the
original percolation model (OPM) and which is not the case in most of
the EP models. An empty lattice site may be a common NN site of more
than one cluster. Since we occupy the empty sites of the smallest
cluster first, the status of occupation or rejection of such sites
cannot be altered in future at the time of growth of the other
clusters of higher sizes if they are encountered as their
neighbors. As soon as all the clusters present at that time are
allowed to grow one layer of NN perimeter sites, MC time step is
increased by one. During a MC step, some of the isolated clusters may
found in contact with each other at the end of one layer
growth. Clusters found in contact are merged together, its size is
relabeled and the cluster size distribution is updated. The growth
probability $g_s(t)$ is re-calculated after every update of cluster
size distribution and second layer of growth starts. The growth
process stops when no empty site on the perimeter of any of the
clusters is available to occupy. In this model, even if all the
clusters in the whole lattice merged to a single large cluster, the
growth of the largest cluster will not be seized. The final
equilibrium cluster size distribution is collected at the end of the
growth process and used to analyze PT.

\begin{figure}[t]
 \centerline{\hfill\psfig{file=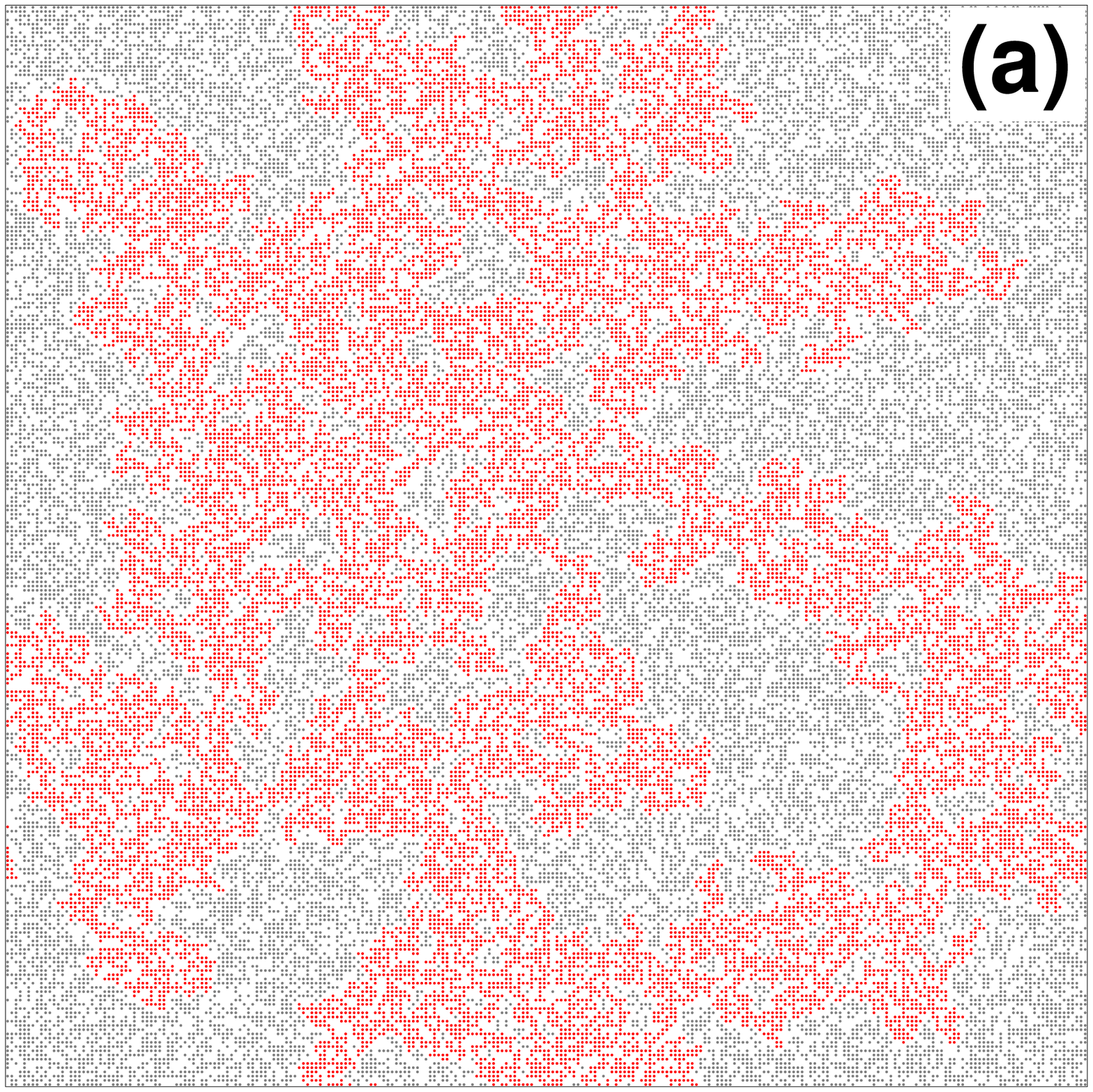, width=0.2\textwidth}
\hfill\psfig{file=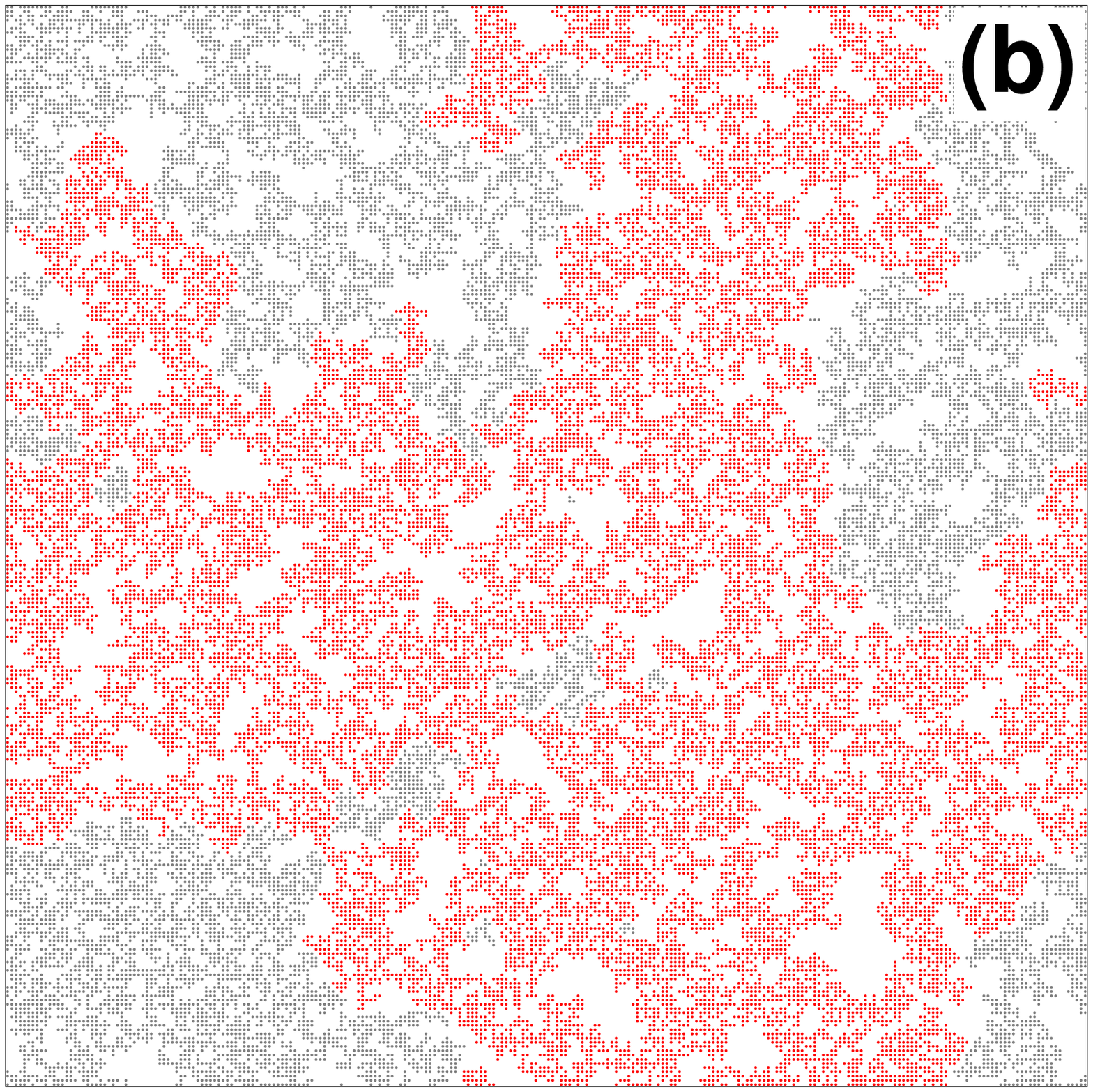, width=0.2\textwidth}
\hfill}
\caption{\label{pic} The final cluster configurations on a square
  lattice of size $L=256$ at $g_{0c}(L)=0.177$ for $\rho=0.50$ (a) and
  at $g_{0c}(L)=0.713$ for $\rho=0.02$ (b). The spanning cluster is
  shown in red. Sites in gray belong to finite clusters. The white
  space represents the inaccessible lattice sites.}
\end{figure}

Extensive computer simulation has been performed for different values
of $L$ from $128$ to $2048$ taking $0.01\le\rho\le 0.59$. Though
$\rho$ has a trivial upper limit, the percolation threshold, the lower
limit should be chosen in such a way that the system has sufficient
number of growth centers. The system is then studied varying $g_0$ for
a given $L$ and $\rho$. Clusters are grown applying periodic boundary
condition (PBC) in both the horizontal and the vertical
directions. Ensemble average is made on $10^5$ random initial
configurations for each $\rho$ and $g_0$ on a given $L$. Typical
cluster configurations for $\rho=0.50$ and $0.02$ at their respective
thresholds $g_{0c}(L)$ on a lattice of $L=256$ are shown in
Fig. \ref{pic}. The morphology of the spanning clusters (in red) are
very different for $\rho$s. For $\rho=0.50$, most of the lattice sites
are found occupied at the end of the growth process, several finite
clusters (in gray) of many different sizes are found within the
spanning cluster as in the OPM. Whereas for $\rho=0.02$, a large
number of lattice sites remain unoccupied as excluded area, almost no
finite cluster is found inside the spanning cluster as it is observed
in percolation of active gels \cite{enclv}. The spanning cluster at a
smaller $\rho$ looks more compact than that at a higher $\rho$.

PT in SCGP is characterized by the properties of the final equilibrium
spanning/large clusters. The order parameter, the probability to find
a lattice site in the spanning cluster, is defined as
$P_{\infty}=S_{max}/L^2$, where $S_{max}$ is the size of the spanning
cluster. The FSS form of $P_{\infty}$ is then expected to be
\begin{equation}
\label{fsspf}
P_{\infty}=L^{-\beta/\nu} \widetilde{P}_{\infty}[(g_0-g_{0c})L^{1/\nu}]
\end{equation} 
where $g_{0c}$ is the critical value of growth parameter at which the
PT occurs. The average value of $S_{max}$ at the threshold scales as
$\langle S_{max}\rangle\approx L^{d_f}$, where $d_f=d-\beta/\nu$ is
the fractal dimension of the spanning cluster. Following the formalism
of analyzing thermal critical phenomena \cite{binder, bruce}, the
distribution of $P_\infty$ is taken as
\begin{equation}
\label{pinfd}
P(P_{\infty})=L^{\beta/\nu} \widetilde{P}[P_{\infty}L^{\beta/\nu}]
\end{equation}
where $\widetilde{P}$ is a universal scaling function. Such a
distribution function of $P_\infty$ is also used in the context of PT
recently \cite{grass}. With such scaling form of $P_\infty$
distribution, one could easily show that $\langle P_{\infty}^2\rangle$
as well as $\langle P_{\infty}\rangle^2$ scale as $\sim
L^{-2\beta/\nu}$. The susceptibility is defined in terms of the
fluctuation in $P_\infty$ as
\begin{equation}
\label{chiinf}
\chi_{\infty}= [\langle S_{max}^2\rangle - \langle
  S_{max}\rangle^2]/L^2.
\end{equation}
Following the hyper-scaling relation $d\nu=\gamma+2\beta$, the FSS
form of $\chi_{\infty}$ is obtained as
\begin{equation}
\label{chiinffss}
\chi_{\infty}= L^{\gamma/\nu} \widetilde{\chi}[(g_0-g_{0c})L^{1/\nu}]
\end{equation}
where $\widetilde{\chi}$ is the scaling function. Studying FSS of
$P_\infty$ and its fluctuation $\chi_{\infty}$, the critical
thresholds $g_{0c}(L)$ are identified and the values of $\beta/\nu$,
$\gamma/\nu$ are estimated. The order of transition is verified by
estimating higher order Binder cumulant (BC)
\cite{kbinder,botet}. Below we present data for two extreme values of
$\rho$, $0.50$ and $0.02$, and we comment on data for the intermediate
range of $\rho$.

\begin{figure}[t]
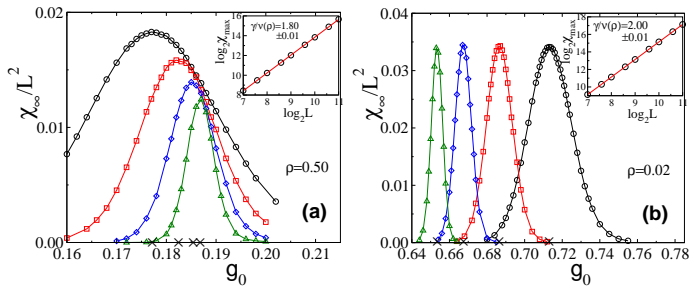

\centerline{\psfig{file=santra_fig2a.eps, width=0.25\textwidth}
 \psfig{file=santra_fig2b.eps, width=0.25\textwidth} }
\caption{\label{xinf} (Color online) Plot of $\chi_{\infty}/L^{2}$
  $vs$ $g_0$ for $\rho=0.50$ (a) and for $\rho=0.02$ (b) on different
  lattice sizes $L= 256[\bigcirc]$, $512[\square]$, $1024[\Diamond]$,
  $2048[\triangle]$. Crosses on the $g_0$ axis represent the
  thresholds $g_{0c}(L)$. In the insets, $\chi_{\rm max}$ is plotted
  against $L$.  } 
\end{figure}

\begin{figure}[!t]
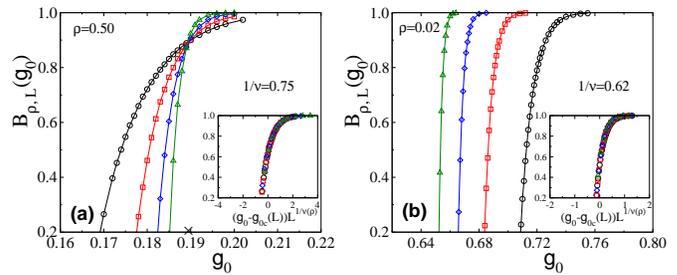

\centerline{
  \psfig{file=santra_fig3a.eps, width=0.24\textwidth}
  \hfill\psfig{file=santra_fig3b.eps, width=0.24\textwidth}}
\caption{\label{binder1} (Color online) Plot of $B_{\rho,L}(g_0)$ $vs$
  $g_0$ for $\rho=0.50$ (a) and for $\rho=0.02$ (b) for different $L$
  using the same symbol set of Fig.\ref{xinf}. In the insets,
  $B_{\rho,L}(g_0)$ is plotted against
  $(g_0-g_{0c}(L))L^{1/\nu(\rho)}$.}
\end{figure}

In Fig. \ref{xinf}, $\chi_{\infty}/L^{2}$ is plotted against $g_0$ for
two different values of $\rho$: $0.50$ (a) and $0.02$ (b). As
expected, $\chi_{\infty}$ is found to have a maximum for a particular
value of $g_0$ for a given $L$. The positions of these maxima
$g_{0c}(L)$ correspond to the percolation thresholds in this model and
are marked by crosses on the $g_0$ axis. For $\rho=0.02$ and $L=2048$,
it is found that $g_{0c}(L)=0.6536(2)$ which is higher than $p_c$ of
OPM, the critical occupation probability for growing the percolation
clusters from a single seed following Leath algorithm \cite{LH}, as it
happens in most of the EP growth models \cite{AP,ziff,tric}. Note
that, the threshold here is a non-trivial finite value in contrast to
the trivial threshold value in SCA \cite{avd}. As in the case of
explosive electric breakdown model \cite{electric}, the values of
$g_{0c}(L)$ are found to decrease with increasing $L$ for
$\rho\lesssim 0.4$. The maximum values of the susceptibility
$\chi_{\rm max}$ are expected to follow a scaling relation $\chi_{\rm
  max} \sim L^{\gamma/\nu}$. Values of $\chi_{\rm max}$ for different
$L$ at their respective $g_{0c}(L)$ are plotted against $L$ in the
insets of Fig. \ref{xinf}(a) and (b) for $\rho=0.50$ and $0.02$
respectively. By linear least square fit to the data points, the
values of $\gamma/\nu$ are extracted. For $\rho=0.50$, it is found
that $\gamma/\nu=1.80\pm 0.01$, that of the OPM ($\approx 1.792$
\cite{stauffer}) within error bar. The value of $\gamma/\nu$ remains
unaltered within $\pm 0.02$ for $\rho\ge 0.45$. Hence, the transitions
for $\rho\ge 0.45$ belong to the same universality class of
OPM. Whereas for $\rho=0.02$, $\gamma/\nu$ is found $2.00\pm 0.01$
\cite{herr1,bfwm} as it occurs in a first order DPT. The value of
$\gamma/\nu \approx 2$ is also found to occur for $\rho\le 0.05$
within error bar. In the intermediate region $0.05<\rho<0.45$, the
value of $\gamma/\nu$ is found to decrease continuously from $2.0$ to
$1.80$ as $\rho$ changes from $0.05$ to $0.45$. Such continuously
varying exponents are also observed in a hybrid PT model
\cite{hyb,prl116}. To confirm the nature of PT in
SCGP, the $4$th order BC
\begin{equation}
\label{bc}
B_{\rho,L}(g_0)=(3/2)[1-\langle S_{max}^4\rangle/(3\langle
  S^2_{max}\rangle^2)]
\end{equation} 
is studied. In Fig. \ref{binder1}, $B_{\rho,L}(g_0)$ is plotted
against $g_0$ for different $L$ for $\rho=0.50$ (a) and $\rho=0.02$
(b). For $\rho=0.50$, the plots of $B_{\rho,L}(g_0)$ for different $L$
cross at a point corresponding to the critical percolation threshold
of SCGP, $g_{0c}(\infty) \approx 0.1895$ as it occurs for a CPT. Such
crossing of BCs are also observed for $\rho\ge 0.45$. For $\rho=0.02$,
however, no such crossing of BCs for different $L$ is found to occur
as expected in first order transitions. Non-crossing of BCs are also
observed for $\rho\le 0.05$. In the intermediate region $0.05< \rho<
0.45$, BCs cross over a range of $g_0$ values indicating no precise
crossover value. The FSS form of BC, $B_{\rho,L}(g_0)=
\widetilde{B}[(g_0-g_{0c}(L))L^{1/\nu(\rho)}]$, where $\widetilde{B}$
is a scaling function is verified in the insets of Fig. \ref{binder1},
plotting BC against $(g_0-g_{0c}(L))L^{1/\nu(\rho)}$. To obtain a
reasonable data collapse, the value of $1/\nu(\rho)$ is tuned manually
to $0.75$ for $\rho=0.50$ and to $0.62$ for $\rho=0.02$ with their
respective values of $\gamma/\nu$. Knowing the values of $\gamma/\nu$
and $1/\nu$, the FSS form of $\chi_{\infty}$ is also verified.

\begin{figure}[t]
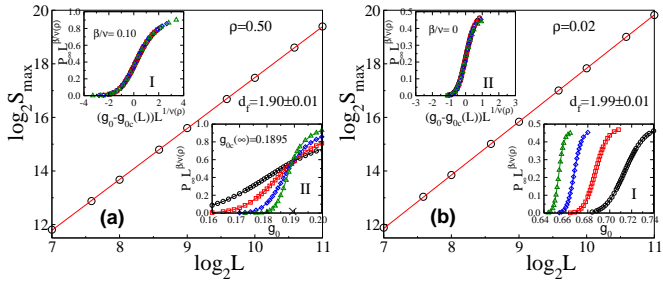

\centerline{
  \hfill\psfig{file=santra_fig4a.eps, width=0.24\textwidth}
  \hfill\psfig{file=santra_fig4b.eps, width=0.24\textwidth}\hfill}
\caption{\label{Pinf} (Color online) Plot of $s_{\rm max}$ against $L$
  for $\rho=0.50$ (a) and for $\rho=0.02$ (b). In the inset-I of each
  plot, $P_{\infty}L^{\beta/\nu(\rho)}$ is plotted against $g_0$. In
  the inset-II of each plot, $P_{\infty}L^{\beta/\nu(\rho)}$ $vs$
  $[g_0-g_{0c}(L)]L^{1/\nu(\rho)}$ is plotted. Same symbol set of
  Fig.\ref{xinf} is used for different $L$.}
\end{figure}

The compactness of the spanning cluster is measured by plotting
$S_{max}\{g_{0c}(L)\}$ against $L$ for $\rho=0.50$ and $0.02$ in
Figs. \ref{Pinf}(a) and (b) respectively. By linear least square fit
through the data points in double logarithmic scale, the fractal
dimensions are obtained as $d_f=1.90\pm 0.01$, as that of OPM, for
$\rho=0.50$ whereas $d_f=1.99\pm 0.01$, as that of space dimension
$d$, for $\rho=0.02$. The values of $d_f$ for both $\rho\ge 0.45$ and
$\rho\le 0.05$ are found to be the expected one within $2\%$
error. For $\rho\le 0.05$, the growth parameter $g_0$ exceeds $p_c$ of
OPM and consequently the initial small finite clusters grow as compact
clusters. The compact spanning cluster at smaller $\rho$ emerges due
to the merging of these compact finite clusters. Such growth of a
macroscopic cluster due to nucleation of small finite clusters is an
essential feature of first order DPT. In the intermediate region
$0.05< \rho< 0.45$, the value of $d_f$ is found changing continuously
with $\rho$. Since $\beta/\nu=d-d_f$, one has $\beta/\nu=0.10$ for
$\rho=0.50$ and $\approx 0$ for $\rho=0.02$. The values of $\beta/\nu$
are verified studying the variation of $P_{\infty}$ against $g_0$. In
the inset-I of Fig. \ref{Pinf}(a) and (b),
$P_{\infty}L^{\beta/\nu(\rho)}$ are plotted against $g_0$ for
$\rho=0.50$ and $0.02$ taking their respective values of
$\beta/\nu$. For $\rho=0.50$, a crossing point (as appeared in the
plots of BC) among the plots for different $L$ is found to appear
whereas for $\rho=0.02$ no such crossing point is observed.
Furthermore, for $\rho=0.02$, $P_{\infty}$ becomes steeper and steeper
as $L$ increases. Such steeper increase in $P_{\infty}$ is also noted
in several other models of EP \cite{bfwm}. The collapse of
$P_{\infty}$ is verified by plotting $P_{\infty}L^{\beta/\nu(\rho)}$
vs $(g_0-g_{0c}(L))L^{1/\nu(\rho)}$ in the inset-II of
Fig. \ref{Pinf}(a) for $\rho=0.50$ and in Fig. \ref{Pinf}(b) for
$\rho=0.02$ taking respective values of $\beta/\nu$ and $1/\nu$. Thus,
at higher $\rho$, $P_{\infty}$ follows usual FSS of OPM whereas at
smaller $\rho$ its scaling becomes independent of $L$. For $0.05<
\rho< 0.45$, unusual scaling of $P_{\infty}$ is found to occur. It is
important to note that no DPT occurs in a similar two parameter model
of random cluster growth without suppression in growing the clusters
\cite{santra}, which represents CPT for the whole range of $\rho$.

\begin{figure}[!t]
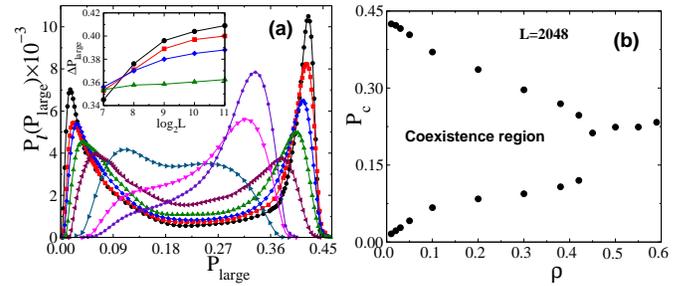

\centerline{
  \psfig{file=santra_fig5a.eps, width=0.24\textwidth}
\psfig{file=santra_fig5b.eps, width=0.24\textwidth}}
\caption{\label{ord}(a) (Color online) Plot of $P_{\ell}(P_{\rm
    large})$ against $P_{\rm large}$ for $L=2048$ for
  $\rho=0.01[\CIRCLE]$(black), $0.02[\blacksquare]$(red),
  $0.03[\blacklozenge]$(blue), $0.05[\blacktriangle]$(green),
  $0.10[\blacktriangleleft]$(maroon), $0.42[\blacktriangleright]$(sky),
  $0.45[\blacktriangledown]$(magenta) and $0.50[\ast]$(violet). In the
  inset of (a), $\Delta P_{\rm large}$ is plotted against $L$ for
  several values of $\rho\le 0.05$. In (b), $P_c$ is plotted against
  $\rho$.}
\end{figure}

To realize the presence of co-existing phases in SCGP, an ensemble of
large clusters, the spanning or the largest if no spanning cluster
appears, at $g_{0c}(L)$ are generated. The probability to find a
lattice site in a largest cluster of size $S_{\rm large}$ is $P_{\rm
  large}=S_{\rm large}/L^2$. The distribution of $P_{\rm large}$ is
expected to be
\begin{equation}
\label{fssop}
P_{\ell}(P_{\rm large})\sim L^{\beta/\nu} \widetilde{P}_{\ell}[P_{\rm
    large}L^{\beta/\nu}]
\end{equation}
where $\widetilde{P}_{\ell}$ is a scaling function. In
Fig. \ref{ord}(a), the distribution $P_{\ell}(P_{\rm large})$,
interpolated through $1000$ equally spaced bins of data points, are
plotted against $P_{\rm large}$ for a wide range of $\rho$. Whenever
there is a crossing point in the BCs, the distributions are obtained
at $g_{0c}$ corresponding to that crossing otherwise they are obtained
at $g_{0c}(L)$. For $\rho\ge 0.45$, not only the distributions are
found single-humped but also the scaled distributions
$P_{\ell}L^{-\beta/\nu}$ collapse onto a single curve when plotted
against $P_{\rm large}L^{\beta/\nu}$. The transitions in this region
are thus CPT which follow usual percolation scaling. On the other
hand, for $\rho\le 0.05$, the distributions are found double-humped
bimodal distributions as it appears in thermal phase transitions
\cite{bruce1} and also reported in some of the EP models
\cite{grass,manna1,tian,marco}. The appearance of bimodal distribution
indicates the coexistence of the spanning cluster with the large
(non-spanning) clusters. No suitable scaling exponent is found to
collapse either of the humps of these bimodal distributions. The
heights of the humps are found increasing with $L$ for a given
$\rho$. Though a leftward shift of the distributions is found to occur
with $L$, the hump to hump separation $\Delta P_{\rm large}$ is found
either constant or increasing with $L$ as shown in the inset of
Fig. \ref{ord}(a) for different values of $\rho\le 0.05$, as in some
of the EP models \cite{electric,manna1}. It is important to note that
$\Delta P_{\rm large}$ is also increasing with decreasing $\rho$ for a
large $L$ making the jump more drastic in the dilute limit of
$\rho$. Not only the compact spanning clusters appear in this region
due to nucleation of finite compact clusters but also the number of
clusters merged to form the spanning cluster become non-extensive with
$L$. All these features provide a strong evidence of true first order
DPT. For $0.05 <\rho <0.45$, $P_{\ell}(P_{\rm large})$ becomes broader
as well as double humps start developing and becomes prominent as
$\rho$ decreases. Thus, in this region, the model exhibits
non-universal critical behavior accompanied by unusual FSS beside a
finite jump in the order parameter like mixed DPT in some of the EP
models \cite{prl116}. However, $\Delta P_{\rm large}$ is found to
decrease with $L$ in the intermediate region of $\rho$. Thus the
apparent DPT in this region may disappear in $L\rightarrow\infty$
limit and one may find the line of CPT is extended further down to
lower values of $\rho$. In Fig. \ref{ord}(b), a phase diagram for SCGP
is presented in the $P_c-\rho$ parameter plane for $L= 2048$ where
$P_c$ is the critical area fraction of spanning (or largest) cluster,
the mean value of $P_{\rm large}$ distribution in the case of a single
hump otherwise the hump positions. Though the boundaries of the
regions are not sharp, it can be seen that a line of second order
transition (for $\rho\ge 0.45$) bifurcates at a tricritical point into
two lines of first order transitions (for $\rho< 0.45$) enclosing a
coexistence region which ultimately represent true first order DPT for
$\rho\le 0.05$. The existence of a tricritical point is also observed in
a growth model with modified product rule incorporating a dilution
parameter \cite{tric}.

In conclusion, an equilibrium SCGP with two parameters is developed
which clearly distinguishes CPT, DPT and mixed DPT in one of its phase
plane. The usual equilibrium spanning cluster approach demonstrates
CPT for $\rho\ge 0.45$ and strong first order DPT for $\rho\le
0.05$. The CPTs are found to belong to the same universality class of
OPM. The DPT, however, is characterized by a discontinuous jump in the
order parameter, coexistence of spanning and non-spanning large
clusters and appearance of compact spanning cluster. A compact
spanning cluster in this region is an outcome of merging of the
compact finite clusters that were grown with high $g_0$. The region of
coexistence is found to be confined within a double-humped bimodal
distribution of the order parameter. In the intermediate range of
$\rho$, the nature of PT still remains inconclusive as characteristic
features of both CPT and DPT appear concurrently and can only be
resolved in true thermodynamic limit.

\end{document}